\documentclass{article}
\usepackage{spconf}
\usepackage{amsmath,graphicx, amssymb}

\usepackage{float}
\usepackage{cite}
\usepackage{mathtools}

\usepackage[font={footnotesize}]{caption}
\usepackage{capt-of}
\usepackage{url}
\usepackage[nolist]{acronym}
\usepackage{amssymb,multirow}
\usepackage{booktabs, bm}
\usepackage{xcolor}

\newacro{AEC}[AEC]{acoustic echo cancellation}
\newacro{ASR}[ASR]{automatic speech recognition}
\newacro{DNN}[DNN]{deep neural network}
\newacro{TTS}[TTS]{text-to-speech}
\newacro{FRR}[FRR]{false reject rate}
\newacro{FAR}[FAR]{false accept rate}
\newacro{IAEC}[IAEC]{implicit acoustic echo cancellation}
\newacro{KWS}[KWS]{keyword spotting}
\newacro{DD}[DD]{device-directedess}
\newacro{SNR}[SNR]{signal-to-noise ratio}
\newacro{FLOPs}[FLOPs]{floating point operations}
\newacro{SIR}[SIR]{signal-to-interference ratio}
\newacro{GSCv2}[GSCv2]{Google speech commands v2}
\newacro{NLMS}[NLMS]{normalized least mean squares}

\title{Implicit Acoustic Echo Cancellation for Keyword Spotting and Device-Directed Speech Detection}
\name{Samuele Cornell$^{1*}$\thanks{*Work performed while at Amazon Alexa.
}, Thomas Balestri$^2$, Thibaud Sénéchal$^2$}
\address{
$^1$Università Politecnica delle Marche, Italy\\
$^2$Amazon Alexa AI, USA 
}

\begin{document}

%
\maketitle

\begin{abstract}
In many speech-enabled human-machine interactions, user speech can overlap with the device playback audio.
In these instances, the performance of tasks such as keyword-spotting (KWS) and device-directed speech detection (DDD) can degrade significantly. 
To address this problem, we propose an implicit acoustic echo cancellation (iAEC) framework where a neural network is trained to exploit the additional information from a reference microphone channel to learn to ignore the interfering signal and improve detection performance. 
We study this framework for the tasks of KWS and DDD on, respectively, an augmented version of Google Speech Commands v2 and a real-world Alexa device dataset. 
Notably, we show a 56\% reduction in false-reject rate for the DDD task during device playback conditions. We also show comparable or superior performance over a strong end-to-end neural echo cancellation baseline for the KWS task with two order of magnitude less computational requirements.
\end{abstract}
\vspace{-0.3cm}
\section{Introduction}
\label{sec:intro}

In the past few years, we have seen a rise in popularity of smart-home devices and virtual assistants. 
Reliable operation of these systems requires addressing many challenging problems \cite{umbach} 
and includes scenarios where user speech can overlap with the device playback audio \cite{takeda2009ica, takeda2012efficient, nishimuta2014development}.  
This phenomenon, sometimes called ``barge in" \cite{takeda2009ica, nishimuta2014development}, happens as well in human-to-human interactions where one
interlocutor begins speaking before another has finished.
Such overlap, is especially problematic in human-device interactions, since usually the \ac{SIR} between the user speech and device playback is low as the device loudspeakers are much closer to the device microphones than the user is.
Moreover, even when user speech and device playback do not overlap, if the playback audio consists of speech, e.g. when it is a \ac{TTS}-generated response to an user query or podcast audio, many tasks become arduous. 
For example, considering custom/multi-KWS \cite{ding2020textual, neuralAECHoward}, without appropriate countermeasures, a model will be prone to pick up also keywords from device \ac{TTS} playback, especially as \ac{TTS} models are increasingly realistic or include celebrity-derived custom voices.
This can lead to the device ``self waking" and continuously interrupting itself as the model, alone, cannot implicitly distinguish between user and device speech and ignore this latter. 
Such problem also affects \ac{ASR} or keyword-less initiated interactions, such as device-directed detection (DDD)\cite{mallidi2018device, huang2019study, gillespieImproveDVAD, norouzianAttDVAD}. 
One trivial way to mitigate this issue would be disabling the KWS functionality while the device is in playback. Yet, doing so prevents the user to ``barge in", making the interaction significantly less natural and intuitive for the user. 

Following \cite{takeda2009ica, takeda2012efficient, nishimuta2014development}, this problem is best formulated in the acoustic echo cancellation (AEC) framework as usually the device playback audio is known. 
We can thus define the playback \emph{reference signal} $\bm{r}$ and a mixed signal $\bm{y} =  \Gamma(\bm{r}) + \bm{u}$, where $\Gamma(\cdot)$ is a (possibly non-linear) function and $\bm{u}$ is the \emph{target signal} for the task at hand i.e. the signal which would be captured by the device if there wasn't any playback return signal $\Gamma(\bm{r})$. This can include instances where there is no user ``barging in", i.e. for which $\bm{u}$ is only background noise, which the classifier should ignore. 
A common model for $\Gamma(\cdot)$ is to use a linear approximation such that $\bm{y} = \bm{r} *\bm{h} + \bm{u}$, where $\bm{h}$ is the impulse response that characterizes the propagation of $\bm{r}$ and includes effects from the room, speaker, and microphone. We will refer to $\bm{n} = \Gamma(\bm{r})$ as the \emph{interferer signal} hereafter. Note that $\bm{u}$ itself could be the user far-field speech with reverberation and other interferers. In this work however we only focus on robustness against the device playback return signal $\bm{n}$. We also restrict ourselves to the monaural case, where the mixture signal is single-channel. 

AEC techniques can be employed to obtain an estimate of the target $\bm{u}$ by leveraging the reference signal $\bm{r}$. These include both classical  \cite{benesty2001advances, hansler2005acoustic, bendersky2008nonlinear, schwarz2013spectral, panda, takeda2009ica, takeda2012efficient, nishimuta2014development} and neural-based (nAEC) methods \cite{zhang2018deep, lei2019deep, fazel2019deep, zhang2019deep,
fazel2020cad,  neuralAECHoward} which are generally more effective. These latter however require the oracle target signal to be available at training time, which is difficult to obtain directly on real-world data, at least in a scalable way. Thus, nAEC methods rely on synthetic data for training, which is inherently mismatched with respect to real-word data.
To counter this mismatch, \cite{neuralAECHoward} proposes the use of an additional \ac{ASR} auxiliary loss obtained from the latent representation of the encoder. Instead \cite{ding2020textual} leverages ASR in inference and proposes a framework for cancelling the TTS playback interferer $\bm{n}$ by using the textual source of the reference TTS signal. 
However, this latter is only applicable to TTS playback and does not generalize to other forms of playback audio such as podcasts or music. In addition, nAEC methods that rely on ASR cannot be used for always-on frontend applications such as KWS since it would be too computationally expensive to perform \ac{ASR} continuously on resource-constrained edge devices.

Focusing on KWS and DDD tasks, we propose to directly feed the reference signal $\bm{r}$ as an additional input to the back-end task model together with the mixture signal captured by the device. We show that such access to the reference signal allows the KWS or DDD classifier to disambiguate between the target and the playback return signal and learn to ignore this latter without the need for an AEC or nAEC pre-processing front-end. 
We call this approach \emph{implicit Acoustic Echo Cancellation} (iAEC).
This method allows for considerable computational resources savings as the computational overhead for having a classifier taking also the reference signal in input is extremely low. We show that competitive performance can be obtained with minimal modifications of the network architecture. We study two different strategies for feeding the reference channel to the back-end classifier, one that involves concatenation and another based on latent-representation masking. We found this latter choice especially promising as it allows for the KWS and DDD architecture to be unchanged (no computational overhead) when there is no playback, which is the predominant scenario in deployment. 

Importantly, in this paper we consider DDD for always-on, streaming scenarios. Previous DDD works \cite{mallidi2018device, huang2019study, gillespieImproveDVAD, norouzianAttDVAD} performed DDD downstream of a KWS model. These systems are not always running, are more resource intensive, and have higher latency than front-end components. Here instead we consider the use of a DDD model that is run continuously and subsumes the role of a KWS model, allowing for a full keyword-free interaction. Understandably, as it is keyword-free and continuously running, such model is especially affected by the device playback issue and prone to the ``self wake" issue. 

We perform our experiments using two datasets: for DDD we use a dataset collected from Alexa assistant devices, while for multi-KWS experiments we instead use a purposely developed augmented version of \ac{GSCv2} \cite{speechcommandsv2} which simulates a smart-home device in both playback and non-playback scenarios.

Finally, as an additional contribution, we devise a data augmentation strategy that helps avoiding bias in training data, boosts performance and allows to train iAEC and nAEC methods even on data with no device playback conditions.

\section{Implicit Acoustic Echo Cancellation}\label{sec:iaec_method}

In this work, we focus strictly only on \ac{KWS} and DDD tasks with the goal of devising a computationally efficient strategy to improve the performance of such tasks during device playback, without incurring in a degradation in normal conditions (non-playback).   

During playback, a KWS or DDD classifier generally observes only the mixture $\bm{y} =  \Gamma(\bm{r}) + \bm{u}$. As said, this presents a challenge when the interferer $\bm{n} = \Gamma(\bm{r})$ and target $\bm{u}$ could lead to an ambiguity for the task at hand (e.g. the interferer $\bm{n}$ contains a keyword but the target $\bm{u}$ does not). In such situations, training a classifier on the mixture signal $\bm{y}$ without access to the reference could bias the model to learn unintended characteristics of the interferer signal. For example, the TTS voice/gender or the fact that the playback interferer has usually more energy than the user speech as the loudspeakers are close to the microphones. These biases could degrade performance severely when the model is deployed.
One way to obviate to this is to feed to the KWS or DDD classifier the reference signal along with the mixture signal ones to aid in the classification task.
The name iAEC stems from the fact that the model here does not have to explicitly recover the target signal as in AEC methods (an arguably more complex task), but only learn how to use the additional reference input $\bm{r}$ to ignore the playback return signal $\bm{n}$ and classify correctly the target signal $\bm{u}$.

In this framework, the goal is to learn a function $\mathcal{F}\left(\bm{y}, \bm{r}, \bm{\theta} \right)$ parametrized by $\bm{\theta}$ that models the joint conditional distribution $P({y}_{\tau}| \mathbf{y}, \mathbf{r})$ where ${y}_{\tau}$ are the labels for the task at hand belonging to the target signal $\bm{u}$ (e.g. for KWS, keyword or non-keyword). Importantly, this formulation includes non-playback conditions, e.g. for which the reference $\bm{r}$ and thus the interferer $\bm{n} = \Gamma(\bm{r})$ is zero and the mixture signal is simply equal to the target $\bm{y} = \bm{u}$. For such instances the problem simply resolves to modeling $P({y}_{\tau} | \mathbf{u})$ as in ``classical" KWS or DDD. 

\subsection{Reference Signal Fusion}\label{sec:iaec_fusion}
\subsubsection{Concatenation: iAEC-C}\label{sec:iaec_c}
Since we focus in this work on KWS and DDD tasks, where usually features like log Mel-filterbank energies (LFBEs) are employed, we can concatenate the mixture and reference signals feature vectors, respectively $\mathbf{x}_y(k)$ and $\mathbf{x}_r(k)$, and feed them to the classifier.
To make notation less cumbersome, we drop the frame index $k$ in the following. This simple method could allow the classifier to exploit the reference channel information. 

As depicted in Figure \ref{fig:mask}, left panel, during playback mode, we apply batch normalization \cite{ioffe2015batch} (BN) separately to the reference and mixture branches prior to concatenation. This is done because reference and mixture signal have usually very different gains, as the latter is usually far-field speech. During non-playback conditions, the input is concatenated with the learned bias parameters $\beta$ of the BN layer. This is equal to concatenating the mean of the post-normalized input features (see left-bottom panel of Figure \ref{fig:mask}). 
\vspace{-0.2cm}
\subsubsection{Masking: iAEC-M}\label{sec:iaec_m}
\begin{figure}
\centering
\includegraphics[width=8.5cm]{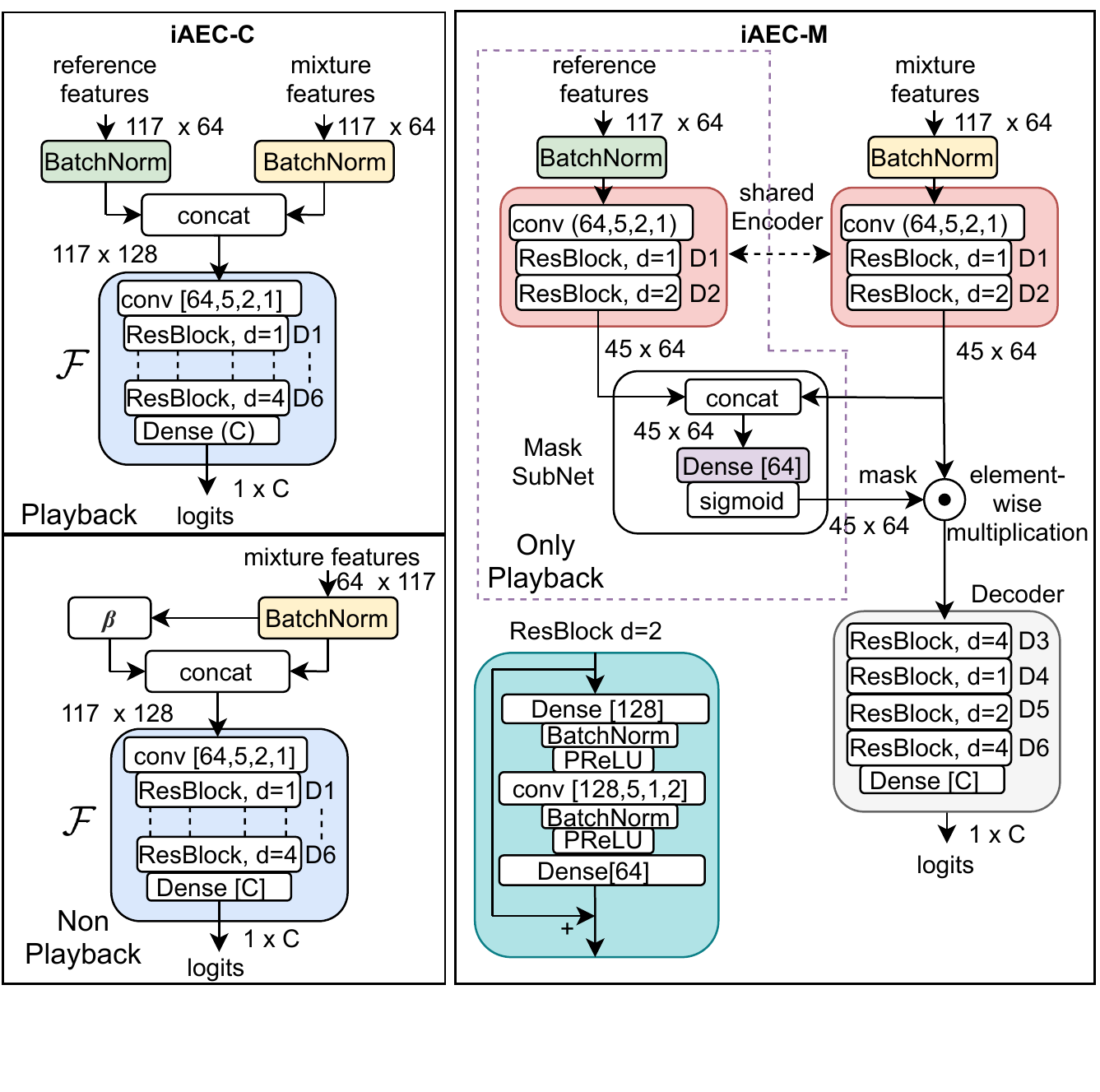}
\vspace{-1cm}
\caption{Schematic of implicit acoustic echo cancellation with concatenation (iAEC-C) and encoder-masking-decoder (iAEC-M) architectures. The LFBE feature extraction part is omitted for simplicity. Neural blocks and input features are described in detail in Section \ref{sec:arcdetails}. Tensor dimensions are represented as \textit{sequenceLength} $\times$ \textit{featureMaps} and we display the values used during training. 1D convolutional blocks (conv) are represented as [\textit{featureMaps}, \textit{kernelSize}, \textit{stride}, \textit{dilation}] while linear layers (Dense) as [\textit{featureMaps}]. 
Regarding iAEC-M, we show the best configuration (D2) as found in Section \ref{sec:ddresult}.
We also report for convenience the layers used in each ResBlock, which has the same structure as in \cite{luo2018convtasnet, cornell2020detecting}.
}
\label{fig:mask}

\end{figure}

\vspace{-0.1cm}
Concatenating the input feature with the learned $\beta$ parameters from the BN layer is a waste of computing resources in non-playback conditions, which account for most of the deployment time. 
This problem can be obviated by using the reference to produce a sigmoid mask which is applied to the latent representation of the mixture signal, as illustrated in the right panel of Figure \ref{fig:mask}.

Compared to iAEC-C, here $\mathcal{F}$ is split into an encoder $\mathcal{E}$ and decoder $\mathcal{D}$. 
The encoder is shared between reference and mixture branches to save L1 cache memory and produces two latent representations: the reference embedding $\bm{Z}_{r} = \mathcal{E}(\bm{x}_r)$ and the mixture embedding $\bm{Z}_{y} = \mathcal{E}(\bm{x}_y)$.

These latent representations are $\in \mathbb{R}^{T \times D}$, where $T$ is the sequence length and $D$ the embedding size. They are concatenated and used to derive a mask through a linear projection layer $\mathcal{P}$ with weight matrix $\in \mathbb{R}^{2D \times D}$, followed by a sigmoid activation $\sigma(\cdot)$: $\bm{M} = \sigma(\mathcal{P}([\bm{Z}_{y}, \bm{Z}_{r}]))$. This mask is then applied to the encoded representation from the mixture branch
$\bm{Z}_{\gamma} = \bm{M} \odot \bm{Z}_{y}$ via element-wise multiplication. This operation acts as a gating mechanism over $\bm{Z}_{y}$. Finally $\bm{Z}_{\gamma}$ is fed to $\mathcal{D}$ to obtain the predictions. 
In non-playback mode the masking mechanism along with the entire reference branch are dropped and $\bm{Z}_{\gamma} = \bm{Z}_{y}$ directly. 

\subsection{On-the-fly Data Augmentation}\label{sec:data_augm}
In some application scenarios the reference playback signal $\bm{r}$ from the device may be not available for training and only the mixture $\bm{y}$ is available. In edge-applications for example, the playback TTS signal is usually not uploaded to the server-side to save bandwidth. Moreover, regarding smart-home devices, indeed most of the examples available in training can feature limited or no user barge in scenarios as e.g. could be an experimental/not fully supported feature. 
In these instances one obvious solution is 
to add artificial device playback via simulation. This may require a complex pipeline involving room, loudspeaker and front-end simulations.

Instead, here, as an additional contribution, we propose a simple but effective on-the-fly data-augmentation strategy to generate ($\bm{y}, \bm{r}, \bm{n})$ triplets by sampling multiple mixture signals $\bm{y}$ from the training set.

Given a collection of $N$ training examples $\textbf{x}_l$ and corresponding labels ${y}_l$ for the task at hand $\left\{(\textbf{x}_l, {y}_l)\right\}_{1\hdots N}$ we randomly sample two examples  $(\textbf{x}_i, {y}_i)$ and $(\textbf{x}_j, {y}_j)$. 
These two examples are arbitrarily assigned the role of target $\bm{u} = \textbf{x}_{i}$ and reference $\bm{r} = \textbf{x}_{j}$ no matter their original labels.
We then generate the mixture $\bm{y}$ by applying a random time-shift to $\textbf{x}_{j}$ and mixing it with $\textbf{x}_{i}$ at a randomly chosen \ac{SIR}. The corresponding mixture $\bm{y}$ is assigned label ${y}_{\tau} = {y}_{i}$ and the interferer label ${y}_j$ is ignored. This strategy forces the model to learn to ignore $\textbf{x}_{j}$ by using the reference $\bm{r}$, whatever the original label $y_{j}$. As $\textbf{x}_{j}$ can contain a keyword from a user, without the reference $\bm{r}$ it would be impossible for the model to ignore the interferer signal (as it is a legit user keyword) and output the correct prediction relative to the target $\bm{u}$.
Because the target and interferer belong effectively to the same distribution, this data augmentation can also mitigate potential bias in the training data and boosts generalization to new speakers as we show in Section \ref{sec:ddresult}.
Note that it is also possible to leverage unlabeled data when assigning the interferer. 

We show that, as far our application is concerned, this naive technique is competitive with simulation via image-method techniques such as gpuRIR \cite{diaz2021gpurir} and, used as an additional data-augmentation strategy, can improve the results even when the true oracle target, interferer and reference are available in training (see Section \ref{sec:kws}). We show in Section \ref{sec:kws} that can also boost performance of nAEC models.






\vspace{-0.3cm}

\section{Datasets}


\subsubsection{Speech Commands Mix}\label{sec:speech_command_mix}

To study how playback can degrade KWS in a controlled scenario, we extended \ac{GSCv2} \cite{speechcommandsv2} with TTS and music. 
We generated two additional versions of the original dataset by mixing TTS from LibriTTS \cite{libritts} and music from Musan \cite{snyder2015musan}, respectively.
To generate challenging TTS interferers, we sampled segments from LibriTTS containing GSCv2 keywords. The temporal location of the keywords from LibriTTS was determined using forced alignment. Interferer and reference pairs are generated using gpuRIR \cite{diaz2021gpurir} by adding artificial reverberation to the original LibriTTS segments. 
For each room impulse response, we sampled a room size from uniform distribution $U(10, 50)$ $m^2$ and T60 reverberation time from $U(0.2, 0.6)$ s. The position of the source is chosen randomly inside the virtual room. To simulate a smart-speaker device, the microphone position is constrained to be in a radius of 5\,cm from the source, oriented outward with cardioid polar pattern. 
Mixture files are obtained by mixing the reverberated interferer signal and original \ac{GSCv2} signal with \ac{SIR} $\sim U(-12, 3)$ dB, which is consistent with what is observed on Alexa devices. An equivalent procedure is followed for mixtures with music. We use the official \ac{GSCv2} training, development and test split in our experiments.
We made the scripts to generate this dataset available\footnote{\url{https://github.com/popcornell/SpeechCommandsMix}}.

\vspace{-0.3cm}
\subsubsection{Alexa Dataset}\label{sec:alexa_dataset}

Amazon Alexa assistant provides a ``follow-up" mode (FUM) \cite{gillespieImproveDVAD} that allows users to interact with the device agent without repeated use of the wakeword. We leverage this data corpus for training and evaluating our DDD model. Ground truth DDD speech annotations are provided for each FUM utterance. The dataset consists of 538k utterances split into train, dev, and test partitions of 465k, 20k, and 53k utterances, respectively.

As FUM does not contain any playback data, in our experiments we also used a TTS model to generate synthetic playback signals for the default Alexa voice. 
In our experiments we also report results when another, unmatched voice (\emph{Matthew}) is used in training. We generated 500k clean TTS utterances and used gpuRIR to add artificial reverberation with the same configuration described in Section \ref{sec:speech_command_mix}. For evaluating our models under TTS playback conditions, we used an Alexa Echo Show 10 device to record both playback-only device responses and instances where 10 speakers were tasked to say commands over the TTS audio.
The default TTS Alexa voice was used. 
This lab collection resulted in a corpus of approximately 2 hours and 45 minutes which was split equally in development and test partitions. 
Note that both training/dev and test sets can contain background environmental noise in the target signal. 


\section{Experiments}

\subsection{Architecture and Training Details}
\label{sec:arcdetails}
 
 For our experiments we employ the Temporal Convolutional Network (TCN) from \cite{cornell2020detecting} with a few modifications. Firstly, we use here a smaller network with $X=3$ residual blocks (ResBlock in Figure \ref{fig:mask}) and $R=2$ repeats for a total of 6 blocks. Secondly, we employ an initial 1D convolutional layer with kernel size 5 and stride 2 instead of a bottleneck convolutional layer. Thirdly, BN is used instead of global layer normalization \cite{luo2018convtasnet} and depth-wise convolutions in ResBlocks have kernel size of 5.
A final linear layer with output size $C$ is used to derive the class logits. A sigmoid activation function is used for DDD ($C=1$) while softmax is used for multi-KWS experiments ($C=35$ posteriors, corresponding to the number of keywords in GSCv2, both original and our augmented version).
The total number of parameters is 131k.
 We use $64$-dimensional LFBEs as input features extracted with a Short-Time Fourier Transform (STFT) $25$ ms window and $10$ ms stride. 
 The model is trained using cross-entropy loss on segments of $117$ frames, which is the receptive field of the model. If the input length is less than $117$ frames, zero padding is employed. 
 During testing, we employ max pooling if the input feature sequence is longer than the receptive field producing a single prediction for the whole sequence. For regularization, we apply SpecAugment \cite{spec_augment} independently on both reference and input mixture LFBE features after the initial BN layers. We use the Adam algorithm \cite{kingma2017adam} for optimization and tune the learning rate, weight decay and SpecAugment hyper-parameters for each experiment using the validation split. Each model is trained for a maximum of $200$ epochs with $10$ epochs early stopping. We use a batch size of $256$ and $1200$ for the multi-KWS and DDD experiments, respectively. All models are trained on both playback (where reference is available) and non-playback conditions.
 
 Since both KWS and DDD models employ LFBEs features in input, in our experiments we perform the proposed on-the-fly data augmentation strategy described in Section \ref{sec:data_augm} in the STFT domain, prior to taking the magnitude and apply the Mel filterbank transform. This leads to a minor training speed-up. The time-shift is applied on STFT frames and is sampled from $U(15, 20)$ frames.
 The interferer and target signals are mixed such that the SIR falls in $U(-20, 3)$dB. 
 Note that such rather extreme frame-wise shifting is necessary to simulate the non-deterministic delay in the device playback and input audio pipeline, mainly due to input/output audio software buffers. This delay leads to substantial misalignment between the reference $\bm{r}$ and corresponding interferer $\bm{n} = \Gamma(\bm{r})$ and must be accounted for during training so the model can learn to compensate for it. 

\subsection{Device-Directed Speech Detection}
\label{sec:ddresult}

In this section we present and discuss our DDD experiments on the real world dataset described in Section \ref{sec:alexa_dataset} focusing on TTS playback conditions. 
We use \ac{FRR} and \ac{FAR} as our metrics and report \ac{FRR} at two fixed \ac{FAR} values (non-playback and playback). The \ac{FAR} values are chosen on the development set according to customer feedback and are redacted in this document due to privacy reasons.
As our goal is to obtain a practical on-device streaming DDD classifier, we
also report the \ac{FLOPs} per output prediction in both playback and non-playback conditions. 

In Table \ref{tab:result_dvad} (upper panel) we report the results for adding simulated interferer signals while training a standard TCN DDD classifier (Baseline) with the architecture explained in Section \ref{sec:arcdetails}. In particular, we compare the strategies of using a test and dev set matched (default Alexa voice) versus unmatched (Matthew voice) TTS model during training.
As expected, adding simulated playback with a matched TTS model (+ Alexa TTS) significantly improves performance especially in playback. 
On the contrary if a unmatched TTS model is used in training, very marginal improvement is observed in playback while in non-playback conditions FRR degrade significantly. 
This suggests that the model is learning to ignore the TTS playback mainly based on the ``identity" and, to a less extent, perceived gender of the TTS speech. The FRR in non-playback increases for the second model because his perceived gender is male, and the test set is composed mainly by male speakers (while the Alexa TTS voice perceived gender is female). This biases the model to be more prone to consider male speakers as TTS and reject them.

In the second panel, we study instead the effect of extending the classifier by simply concatenating the reference channel features at the first layer, after BN, as explained in Section \ref{sec:iaec_c}. This model requires slightly more FLOPs than the standard Baseline TCN classifier.
As our training dataset lacks playback conditions (see Section \ref{sec:alexa_dataset}) also here we resort to simulation using both matched and unmatched TTS voices. However we also study the effect of the data augmentation strategy outlined in Section \ref{sec:data_augm} which can be leveraged now since the model (iAEC-C) has access to the reference. 

We observe that the proposed simple data augmentation strategy alone (iAEC-C augm) is able to improve FRR in both conditions despite the model is trained with no TTS data but only with legit user queries used both for targets and playback roles. 
Adding simulated matched and, to a less extent, even unmatched TTS interferer/reference data further improves the performance in both conditions. 
In the Matthew TTS case, the proposed method helps fighting potential biases in the training data and prevent overfitting one particular voice/gender.
%

On the other hand if the iAEC-C model is trained only with matched TTS simulated data and no data augmentation strategy (- \emph{augm} + Alexa TTS) we observe a degradation in performance. This hints that the model with reference access is more likely to overfit the simulated interferer acoustic characteristics in the training set and not generalize well to real-world environments despite our simulation efforts. More complex simulation pipelines may be able to mitigate this issue but have their drawbacks (e.g. lots of hand-tuning). 
As smart-home assistants offer more voice options, using the proposed data augmentation allows us to avoid retraining the DDD classifier for each new TTS voice and reduces the risk of rejecting speakers whose voices are similar to the TTS model.



\vspace{-0.2cm}
\begin{table}[!htbp]
	\centering
	\setlength{\tabcolsep}{2pt} 
	\renewcommand{\arraystretch}{1} 
	\footnotesize
	\caption{\ac{FRR} at fixed FAR (redacted) for non-playback and playback (TTS playing) conditions. We study different training strategies using the default Alexa voice and an alternative TTS voice named Matthew. Intra-dataset mixing (intramix) refers to the on-the-fly mixing strategy outlined in Section \ref{sec:iaec_method}.}

	\begin{tabular}{|cl|cc|cc|}
		\hline
		\multicolumn{2}{|c|}{{ \bf Model}} & \multicolumn{2}{c}{{ \bf Non-Playback}} &  \multicolumn{2}{c|}{{ \bf Playback}} \\
		
		& & \ac{FRR}@\ac{FAR} & FLOPs & FRR@FAR & FLOPs \\
		\hline
		\multirow{2}{*}{} & Baseline & 0.189 & 242k & 0.348 & 242k \\
		& + Alexa TTS & 0.187 & -- & 0.241 & -- \\
		& + Matthew TTS & 0.202 & -- & 0.339 & -- \\
		\hline
		
		& iAEC-C (augm) & 0.188 & 283k & 0.258 & 283k \\
		& + Alexa TTS  & \textbf{0.185} & -- & \textbf{0.227} & -- \\
	    & + Matthew TTS & 0.187 & -- & 0.256 & -- \\
	    & - augm +  Alexa TTS  & 0.193 & -- & 0.299 & -- \\
	    
	    \hline
	\end{tabular}
	
	\label{tab:result_dvad}
\end{table}



%



The top panel of Table \ref{tab:result_dvad_cat2} investigates the effect of concatenating at deeper residual blocks for iAEC-C from the 1st (D1, same as in Table \ref{tab:result_dvad}) to 3rd (D3) blocks (see Figure \ref{fig:mask}). All models are trained using the data-augmentation (\emph{augm}) strategy outlined in Section \ref{sec:iaec_method} with no simulated playback TTS data. 
Performance improves with deeper layers up to D2, as the encoder receptive field (FOV) surpasses the maximum offset between reference and interferer signals observed on real-world collected audio, around 200\,ms, as said, mainly due to the output and input software audio buffers. On the other hand, also computation increases with the depth at which concatenation is performed. We can see that concatenation at D2 provides the best trade-off between playback FRR, non-playback FRR, and FLOPs.

\begin{table}[!htbp]
	\centering
	\setlength{\tabcolsep}{3pt} 
	\renewcommand{\arraystretch}{1} 
	\footnotesize
	\caption{\ac{FRR} at fixed FAR (redacted) using iAEC-M and iAEC-C with concatenation at different layers in the TCN model.}

	\begin{tabular}{|cl|c|cc|cc|}
		\hline
		\multicolumn{2}{|c|}{{ \bf Model}} & & \multicolumn{2}{c}{{ \bf Non-Playback}} &  \multicolumn{2}{c|}{{ \bf Playback}} \\
		& & FOV &  \ac{FRR}@\ac{FAR} & FLOPs & FRR@FAR & FLOPs \\
		\hline
		& iAEC-C (input) & 5 & 0.188 & 283k & 0.258 & 283k \\
		& iAEC-C (D1) & 12 & 0.183 & 336k  & 0.178 & 336k  \\
		& iAEC-C (D2) & 28  & 0.184 & 369k & 0.165 & 369k \\
		& iAEC-C (D3) & 60 & 0.183 & 402k & 0.168 &  402k \\
		\hline
		& iAEC-M (D2) & 28 & \bf{0.181} & 242k & \bf{0.150} & 367k \\
	    \hline
	\end{tabular}

	\label{tab:result_dvad_cat2}

\end{table}

The bottom panel of Table \ref{tab:result_dvad_cat2} presents the mask-based approach described in Section \ref{sec:iaec_m}. Masking is applied at the second residual block (D2) as in iAEC-C and shows a further performance improvement on both playback and non-playback conditions. Regarding computational efficiency at D2, iAEC-M requires slightly more \ac{FLOPs} in playback mode than iAEC-C, but significantly less \ac{FLOPs} in non-playback mode, as the reference branch is dropped and the architecture becomes equal to a standard classifier. 
Since this model is ran continuously, playback conditions account for a very small fraction of total inference time and thus iAEC-M leads to the best performance/FLOPs trade-off. 


\subsection{Multi Keyword Spotting}
\label{sec:kws}
In Table \ref{tab:result_kws} we report our results on the augmented GSCv2 dataset described in Sec \ref{sec:speech_command_mix}. 
We report keyword-detection accuracy over the 35 possible GSCv2 keywords for the $3$ different test sets in our augmented version: original GSCv2 (\emph{Non-Playback}), mixed with music (\emph{Playback Music}) and with TTS (\emph{Playback TTS}). 

As with the DDD experiments, we use a standard TCN classifier without access to the reference as our \emph{Baseline}. We also consider a joint model (\emph{+nAEC}) comprised of the state-of-the-art neural AEC model from \cite{neuralAECHoward}, also designed for edge-devices applications, and the \emph{Baseline} TCN classifier: the output of the nAEC is directly fed to the classifier in cascade. 
We also compare with more ``classical" approaches, reporting results for the \emph{Baseline} classifier trained on signals obtained from an oracle (non-causal) Wiener filter with 512 taps and on signals filtered with an STFT-based \ac{NLMS} \ac{AEC} algorithm with 32 taps, step size $\mu=0.5$, 512 and 128 STFT window and hop respectively with square-root Hann window. This latter was implemented using Pyroomacoustics \cite{scheibler2018pyroomacoustics}. 
The oracle Wiener filter results can be regarded as a reasonable upper bound reachable by conventional linear filtering methods. We adapted the implementation from \cite{boeddeker2021convolutive}, we do not report \ac{FLOPs} for this configuration during playback as it is non-causal: \ac{FLOPs} depend on the oracle target signal length. 

For this system we use a weighted loss consisting of a spectral loss term as in \cite{neuralAECHoward} and a KWS loss term instead of an ASR on as in \cite{neuralAECHoward}. The two cascaded models nAEC and TCN are then jointly trained. 
As another baseline, we include the performance for MatchBoxNet-3x2x64 \cite{majumdar2020matchboxnet} (MatchBN) using the official implementation from NeMo toolkit \cite{nemo}.
We also report results for three different training strategies for the models with access to the reference (thus including \cite{neuralAECHoward}). In \emph{orcl} we train the model with access to the oracle reference, interferer and target signals, e.g. the nAEC model is trained to estimate the target and is given in input the reference corresponding to the interferer together with the mixture signal. 
This is a best-case scenario and is only possible because this dataset is fully synthetic. Here there is no mismatch between training and test data, a rather ideal condition which e.g. will prevent the degradation observed in DDD experiments in Table \ref{tab:result_dvad} last row due to mismatched simulated training and real-world interferer acoustic conditions. 

The \emph{augm} denotes the augmentation strategy described in Section \ref{sec:data_augm} where, in training, we generate fake playback mixture signals by mixing original GSCv2 examples. Each example is randomly given the role of interferer/reference or target, without adding any simulated playback TTS or music. 
Finally the strategy denoted \emph{both} denotes a combination of these two: in training one example is either generated on-the-fly with \emph{augm} or comes from \emph{orcl} with 50\% probability.

As expected, we can observe that for both \emph{Baseline} and \emph{MatchBN} playback performance degrades significantly, especially during TTS playback conditions, which can confuse the model. 
Regarding the models with access to the reference (Baseline +nAEC, iAEC-C and iAEC-M), we can see that, in the best case scenario of matched training and testing (\emph{orcl}), all models significantly improve performance over the Baseline classifier especially in playback conditions. The improvement in non-playback is due to the fact that the MatchBN and Baseline models are prone to reject a valid keyword as TTS playback, as they have no access to the reference. 
If only the \emph{augm} strategy  is used in training the performance degrades compared to the ideal \emph{orcl} data but still affords improvement over models with no reference access. On the other hand, as explained, in many real-world applications the reference signals can be difficult to obtain or may be biased (e.g. all examples with playback belongs to one TTS model).
Combining both strategies yields the best performance overall for all models (including nAEC) even surpassing \emph{orcl}. 
Regarding classical digital signal processing applications, we can see that they are generally very effective with the music interferer but, even in the oracle Wiener filter case, struggle with the TTS interferer. On the other hand, the NLMS AEC solution is extremely light computationally, and future works could explore the combination of the proposed approach with such conventional AEC algorithms.

Overall, our proposed iAEC approaches perform competitively with \emph{+nAEC} but uses two order of magnitude fewer FLOPs for each prediction. This makes the proposed methods more suitable for always-on low-resource applications.

\begin{table}[!htbp]
 
	\centering
	\setlength{\tabcolsep}{0.7pt} 
	\renewcommand{\arraystretch}{1} 
	\footnotesize
		\caption{Accuracy and \ac{FLOPs} on GSCv2 for different models and training strategies (see Sec \ref{sec:kws}). Metrics are reported both for original data (\emph{Non-Playback}) and our simulated playback corpus, separately for \emph{TTS} and \emph{Music} playback conditions.}
	
	\begin{tabular}{|ll|cc|ccc|}
		\hline
		\multicolumn{2}{|c|}{{ \bf Model}} & \multicolumn{2}{c}{{ \bf Non-Playback}} &  \multicolumn{3}{c|}{{ \bf Playback}} \\
		& & Acc \% & FLOPs & Acc \%  (Music) & Acc \%  (TTS) & FLOPs \\
		\hline
		MatchBN \cite{majumdar2020matchboxnet} & & 94.47 & 185k & 61.84 & 36.78   & 185k \\
		Baseline & & 93.35 & 242k & 75.01 & 61.71 & 242k \\
		\hline
		+ oracle Wiener & & 94.20 & 242k & 79.52 & 71.52 & n.a. \\
		\hline
		+ NLMS AEC  & & 93.77 & 242k & 78.95  & 63.13  & 243k \\
		\hline
		\multirow{3}{*}{+nAEC} & orcl & 94.06 & 242k & 82.87 & 82.75 & 15M \\
		 & augm & 93.82  & - & 75.04  & 72.21 &  - \\
		  & both & 94.46 & - & 83.55  & \textbf{83.81} &  - \\
		\hline
		\multirow{3}{*}{iAEC-C} & orcl & 94.52 & 283k & 82.60  & 80.79  &  283k \\
		 & augm & 94.56 & - & 77.91  & 77.52 &  - \\
		  & both & 94.74 & - & 83.54  & 82.93 &  - \\
		\multirow{3}{*}{iAEC-M} & orcl & 94.67 & 242k & 83.87 & 82.47 &  367k \\
		 &  augm & 94.49 & - & 78.21  & 77.01 &  - \\
		  & both & \textbf{94.97} & - & \textbf{84.22} & 83.79 &  - \\
		\hline
	\end{tabular}

	\label{tab:result_kws}

\end{table}



\vspace{-0.5cm}
\section{Conclusions \& Future Work}
In this work, we address the problem of boosting KWS and DDD classifiers performance on edge-devices during device playback without degrading non-playback performance. In such scenarios, the playback signal is usually known. We propose a simple and computationally efficient implicit acoustic echo cancellation (iAEC) framework where a classifier leverages the known playback signal (reference signal) to ignore the return ``echoed" playback signal (interferer signal) captured by the device microphones together with the user speech. 

We explored two strategies for feeding the reference signal to our models and found the use of a latent-space masking approach particularly suited for our applications. 
We performed several experiments for the KWS and DDD tasks and found that the proposed method is able to significantly improve the performance for both tasks in device playback conditions. We show on the KWS task that the proposed method obtains comparable performance with a state-of-the-art neural AEC method but has significantly less computational requirements. 
As an additional contribution, we devised an effective data augmentation strategy that is able to further boost performance of nAEC and iAEC models, allows to train such models on examples with no playback interferer and helps reducing bias in the training data. 

Future work could explore other tasks such as ASR and scenarios beyond device playback where oracle double-talk detection is not available.

\bibliographystyle{IEEEbib}
\footnotesize{\bibliography{strings,refs}}

\end{document}